\documentclass[aps,groupedaddress,reprint,showpacs,floatfix,pdftex]{revtex4-2}
\usepackage[whole]{bxcjkjatype}
\usepackage{graphicx}
\usepackage{comment}
\usepackage{tabularx}
\usepackage{dcolumn}
\usepackage{bm}
\usepackage{amsmath}
\usepackage{amssymb}
\usepackage{txfonts}
\usepackage{url}
\usepackage{xcolor}
\usepackage{multirow}
\usepackage{ulem}
\usepackage{cancel}
\usepackage{upgreek}
\newcommand{\tadd}[1]{#1}
\newcommand{\nuoxy}   {$\nu_{\mathrm{O}}$}
\newcommand{\po}      {$P_{\mathrm{O_2}}$}

\newcommand{\ghost}[1]{}
\begin{document}
\title{Control of Impurity Phase Segregation in a PdCrO$_2$/CuCrO$_2$ Heterostructure}
%%%%%%%%%%%%%%%%%%%%%%%%%%%%%%%%%%%%%%%%%

\thanks{This manuscript has been authored by UT-Battelle, LLC, under contract DE-AC05-00OR22725 with the US Department of Energy (DOE). The US government retains and the publisher, by accepting the article for publication, acknowledges that the US government retains a nonexclusive, paid-up, irrevocable, worldwide license to publish or reproduce the published form of this manuscript, or allow others to do so, for US government purposes. DOE will provide public access to these results of federally sponsored research in accordance with the DOE Public Access Plan (http://energy.gov/downloads/doe-public-access-plan).}
\author{Tom Ichibha$^{1}$}
\email[]{ichibha@icloud.com}
\author{Sangmoon Yoon$^{1,2}$}
\author{Jong Mok Ok$^{1,3}$}
\author{Mina Yoon$^{1}$}
\author{Ho Nyung Lee$^{1}$}
\author{Fernando A. Reboredo$^{1}$}
\email[]{reboredofa@ornl.gov}
%%%%%%%%%%%%%%%%%%%%%%%%%%%%%%%%%%%%%%%%%
\affiliation{
  $^{1}$Materials Science and Technology Division, Oak Ridge National Laboratory, Oak Ridge, TN 37831, USA
}
\affiliation{$^{2}$Department of Physics, Gachon University, Seongnam 13306, Republic of Korea}
\affiliation{$^{3}$Department of Physics, Pusan National University, Pusan 46241, Republic of Korea}

%%%%%%%%%%%%%%%%%%%%%%%%%%%%%%%%%%%%%%%%%
\date{\today}
\begin{abstract}
  PdCrO$_2$ films are synthesized on CuCrO$_2$ buffer layer\tadd{s} 
  on Al$_2$O$_3$ substrate\tadd{s}. 
  This synthesis is accompanied by impurity phase segregation, 
  which hampers the \tadd{synthesis} of high quality PdCrO$_2$ films. 
  The potential causes of impurity phase segregation were studied by 
  using a combination of experiments and ab initio calculations.
  X-ray diffraction and scanning transmission electron microscopy experiments
  revealed impurity phases of Cu$_x$Pd$_{1-x}$ alloy and chromium oxides,
  Cr$_2$O$_3$ and Cr$_3$O$_4$, \tadd{in} PdCrO$_2$.
  Calculations determined that 
  \tadd{oxygen deficiency}
  can cause the impurity phase segregation. 
  Therefore, preventing oxygen release from delafossites
  could suppress the impurity phase segregation.
  The amounts of Cr$_2$O$_3$ and 
  Cr$_3$O$_4$ depend differently on temperature and oxygen partial pressure. 
  A reasonable theory-based explanation for this experimental observation is provided. 
\end{abstract}
\maketitle
%%%%%%%%%%%%%%%%%%%%%%%%%%%%%%%%%%%%%%%%%
\section{Introduction}
\label{sec.intro}\ghost{sec.intro}
%%%%%%%%%%%%%%%%%%%%%%%%%%%%%%%%%%%%%%%%%
Delafossites are intriguing materials that can combine 2D electronic conductivity 
in the cation $A$ layers and magnetism in slightly distorted octahedra in the $B$O$_6$ layers, 
which stack alternately \cite{1971RDS_CTP,1971RDS_DBR,1971RDS_JLG}. 
The abundant possible choices of monovalent $A$ and trivalent $B$ cations
lead to a number of delafossite materials
with diverse physical properties \cite{2017APM,2015HNA}. 
The $AB$O$_2$ delafossites were first reported in 1971 by a group of the DuPont Experimental Station
\cite{1971RDS_CTP,1971RDS_DBR,1971RDS_JLG,2017APM}. 
A quarter of century after delafossites were first reported, they received renewed attention
when the transparent $p$-type semiconductor CuAlO$_2$ was discovered
\cite{1997HK_HH,2000HY_NH}.
Simultaneously, Tanaka et al.\cite{1996MT_HT} reported the strong anisotropy of electronic conduction
for the metallic PdCoO$_2$ single crystals \cite{1971RDS_JLG}. 
One decade later, Takatsu and Maeno et al., working on PdCoO$_2$ and PdCrO$_2$,
reported the growth of single crystals of
PdCrO$_2$ \cite{2010HT_YMb}.
These single crystals exhibit intriguing phenomena\cite{2017APM} 
such as the unconventional anomalous Hall effect in PdCrO$_2$\cite{2010HT_YMc} 
and anomalous temperature dependence of specific heat and electrical resistivity
that are driven by high-frequency phonons in PdCoO$_2$\cite{2010HT_YMc}.
Their seminal work originated the continuous study of delafossite metals to this day. 

\vspace{2mm}
Delafossite metals have electronic conductivity comparable with 
the most conductive pure metals
\cite{1971RDS_JLG,1997MT_HT,1996MT_HT,2017APM} 
owing to their remarkably long electronic mean free paths of up to 20 $\upmu$m
\cite{2017APM,2012CWH_EAY,2015PK_PDCK}. 
Among delafossite metals, PdCrO$_2$ is especially interesting
because it coexists with a layerwise non-collinear spin state
\cite{1986JPD_PH,1995MM_HY,2009HT_YM,2014HT_CB,2015DB_HT}
and exhibits high electronic conductivity \cite{2009HT_YM}.
Its topological properties, primarily caused by spin--orbit coupling in Pd,
allow for the observation of an unconventional anomalous Hall effect
\cite{2010HT_YMc,2013JMO_JSK} in bulk PdCrO$_2$.
Additionally, PdCrO$_2$ films and surfaces have been studied.
Angle-resolved photoemission spectroscopy experiments showed that
Pd-terminated PdCrO$_2$ has surface ferromagnetism, which may 
\tadd{originate from the} Stoner-like instability\cite{2018FM_PDCK,2021TH}.
Experimental studies of PdCrO$_2$ films established that the antiferromagnetic
spin state remains stable down to a thickness of 3.6 nm \cite{2020JMO_HNLa}.

\vspace{2mm}
Hybrid layered heterostructures, composed of PdCrO$_2$ and other delafossite materials,
could exhibit interesting and different phenomena than their parent compounds \cite{2020FL_RR}.
However, despite the interest in the material, 
the epitaxial growth of PdCrO$_2$ films has not been widely studied
\cite{2018TH_AT,2019JS_DGS,2019MB_SO,2019PY_HUH,2020JMO_HNLa}.
The growth of PdCrO$_2$ films on Al$_2$O$_3$ is sometimes accompanied
  by impurity phases (i.e., Cu$_x$Pd$_{1-x}$ alloy and chromium oxides) \cite{2020JMO_HNLa}. 
Recent research discovered that a one-monolayer buffer layer of CuCrO$_2$ on an Al$_2$O$_3$
substrate suppresses this instability \cite{2020JMO_HNLa}. 
However, a nonnegligible amount of impurity phase is still formed.
Understanding the mechanism of the impurity phase segregation and how to suppress it
is highly desired for the growth of heterostructures containing PdCrO$_2$ or other Pd-based delafossites. 

\vspace{2mm}
In this work, the mechanism of impurity phase segregation 
of \tadd{a} heterostructure of a PdCrO$_2$ layer with a CuCrO$_2$ buffer layer
on an Al$_2$O$_3$ substrate was studied using a combination of experiments and ab initio calculations. 
X-ray diffraction (XRD) and scanning transmission electron microscopy (STEM) experiments were performed,
and the segregation of Cu$_x$Pd$_{1-x}$ alloy and chromium oxide
(Cr$_2$O$_3$ and Cr$_3$O$_4$) impurity phases was observed. 
These experiments revealed that the  formation of Cr$_2$O$_3$ negatively correlates with oxygen partial pressure, 
whereas the formation of Cr$_3$O$_4$ does not correlate with oxygen partial pressure. 
Moreover, the Cr$_2$O$_3$ (Cr$_3$O$_4$) formation weakly (strongly)
positively correlates with temperature.
The segregation of Cu$_x$Pd$_{1-x}$ alloy and chromium oxide impurity phases must be 
accompanied by the appearance or disappearance of point defects because the segregation processes are not stoichiometric. 
In this scenario, calculations revealed that
oxygen vacancies can cause the impurity phase segregation. 
Calculations also revealed that the segregation of Cr$_2$O$_3$ or Cr$_3$O$_4$ is energetically the most favorable
among the chromium oxides, agreeing with the
experiments described in Section \ref{sec.segre}. 
Finally, the calculations also revealed that the formation of Cr$_2$O$_3$ and Cr$_3$O$_4$ depends on temperature
and oxygen partial pressure. 

%%%%%%%%%%%%%%%%%%%%%%%%%%%%%%%%%%%%%%%%%%%%%%%%
\section{Experimental and calculation details\label{sec.methods}}
\ghost{sec.methods}
%%%%%%%%%%%%%%%%%%%%%%%%%%%%%%%%%%%%%%%%%%%%%%%%
%%%%%%%%%%%%%%%%%%%%%%%%%%%%%%%%%%%%%%%%%
\subsection{Experimental details\label{subsec.expt}}
\ghost{subsec.expt}
%%%%%%%%%%%%%%%%%%%%%%%%%%%%%%%%%%%%%%%%%
A PdCrO$_2$ layer with thickness of approximately 10~nm was grown on a one-monolayer ($\sim$0.38~nm) CuCrO$_2$ buffer layer on an Al$_2$O$_3$ substrate 
via pulsed laser deposition using polycrystalline targets.
Before the film growth, commercially available Al$_2$O$_3$ (0001) substrates (CrysTec, Germany)
were annealed at 1100~$^\circ$C for 1~h to achieve atomically flat surfaces with step-terrace structure.
For PdCrO$_2$ films, the growth conditions were widely varied: temperature ($T$) was 500--800~$^\circ$C, and oxygen partial pressure ({\po}) was 10--500~mTorr.
The repetition rate and fluence of KrF excimer laser ($\uplambda$~=~248~nm) were fixed at 5~Hz and 1.5~J/cm$^2$, respectively.
The cross-sectional STEM specimens were prepared using low-energy ion milling at LN$_2$
temperature after mechanical polishing.
High-angle annular dark field (HAADF) STEM measurements were performed on a Nion UltraSTEM200 operated at 200~kV.
The microscope is equipped with a cold-field emission gun and a third- and fifth-order aberration corrector for sub-angstrom resolution.
The convergence half-angle of 30~mrad was used, and the inner angle of the HAADF STEM was approximately 65~mrad.

%%%%%%%%%%%%%%%%%%%%%%%%%%%%%%%%%%%%%%%%%
\subsection{Calculation details}
\label{sec.calc}\ghost{sec.calc}
%%%%%%%%%%%%%%%%%%%%%%%%%%%%%%%%%%%%%%%%%
Density functional theory (DFT) implemented in the VASP package \cite{1996KRE} was used  
to understand the energetics of competing phases during the experimental growth process.
The Perdew--Burke--Ernzerhof (PBE)+$U$ method \cite{PBE,LDA+U} was used.  
The Hubbard $U$ correction  was applied to the 3$d$ shell of the Cr atoms. 
The $U$ value was 3.3~eV, which was optimized compared with the results of the HSE06 functional \cite{2003JH_ME}, 
as described in the Supporting Information.
The core electrons were replaced with pseudopotentials made by the projector-augmented wave method accompanied by the VASP code 
\cite{1994PEB,1999GK_DJ,2000DH_JH}. 
The cutoff energy was 520~eV, and $k$-spacing was 0.30~\AA$^{-1}$,
which converged the Cr vacancy formation energy in CuCrO$_2$ within 2~meV. 
Experimental lattice vectors for CuCrO$_2$ \cite{2011MF_GE}, PdCrO$_2$ \cite{2018MDL_JGP},
and Al$_2$O$_3$ were used \footnote{The room temperature structure of Ref. \cite{2008SK_NI} was used}.
The lattice vectors reported in the Materials project \cite{2013AJ_KP} were used for chromium oxides and chromium metal 
\footnote{The database ID for each material: Cr = mp-90, CrO = mp-19091, Cr$_3$O$_4$ = mp-756253, and Cr$_2$O$_3$ = mp-19399.}.
The atomic coordinates were relaxed for the functional.
The convergence criteria for the self-consistent field and ionic cycles were 
$1.0 \times 10^{-7}$~eV and $1.0 \times 10^{-6}$~eV, respectively. 

%%%%%%%%%%%%%%%%%%%%%%%%%%%%%%%%%%%%%%%%%
\section{Results and Discussions} 
\label{sec.result}\ghost{sec.result} 
%%%%%%%%%%%%%%%%%%%%%%%%%%%%%%%%%%%%%%%%%
%%%%%%%%%%%%%%%%%%%%%%%%%%%%%%%%%%%%%%%%%
\subsection{Segregation of impurity phases}
\label{sec.segre}
\begin{figure}[htbp]
  \centering
  \includegraphics[width=\hsize]{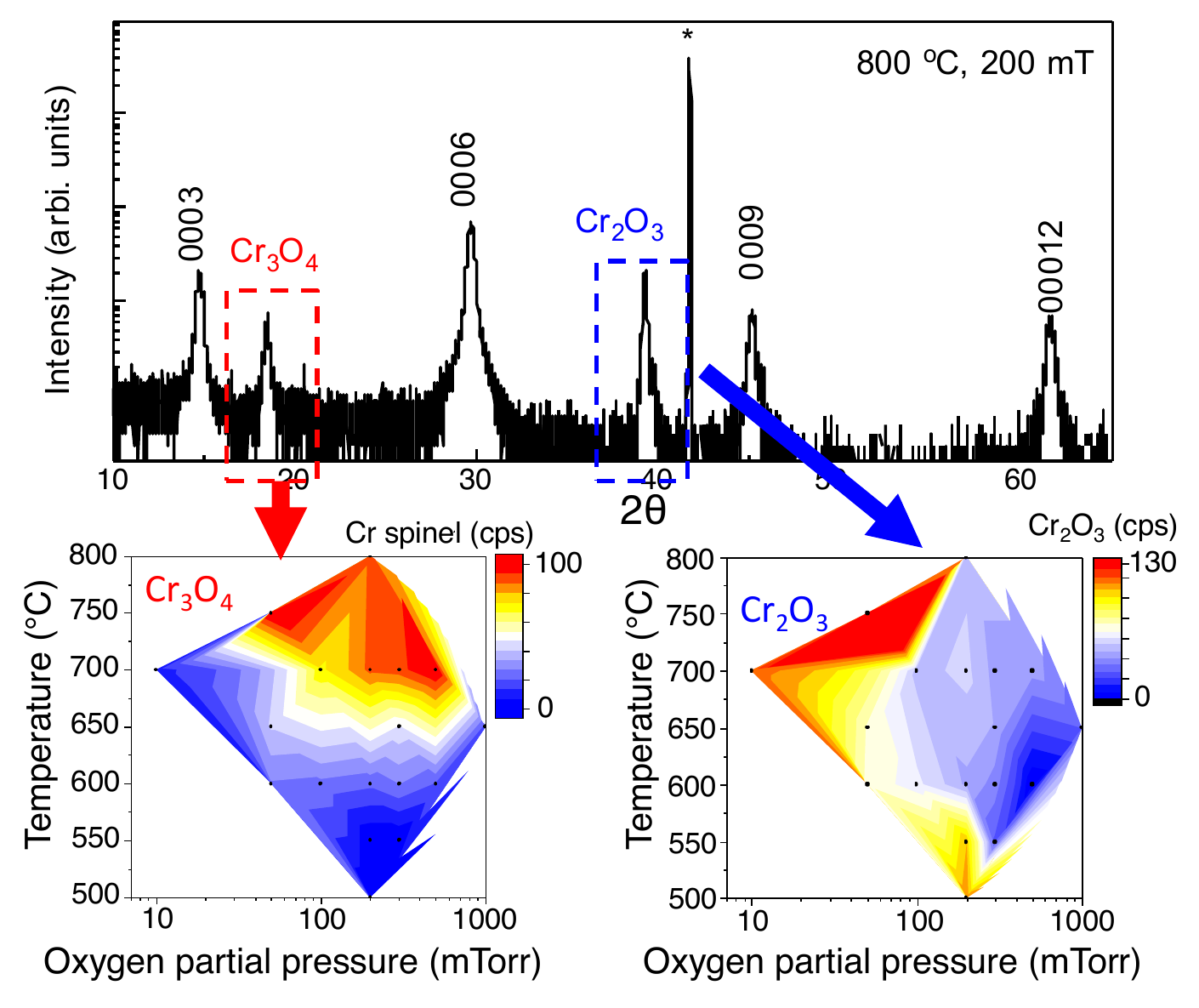} 
  \caption{\label{fig.xray}\ghost{fig.xray} 
    XRD spectrum and mapping of XRD peak strengths of 
    Cr$_3$O$_4$ and Cr$_2$O$_3$ for temperature 
    and oxygen partial pressure. 
    This measurement was performed for an approximately 6 nm PdCrO$_2$
    film grown on a one-monolayer CuCrO$_2$ buffer layer.
  }
\end{figure}

\ghost{sec.segre}
%%%%%%%%%%%%%%%%%%%%%%%%%%%%%%%%%%%%%%%%%
\tadd{
  Impurity phases including Cu$_{x}$Pd$_{1-x}$, Cr$_2$O$_3$, and Cr$_3$O$_4$  
  have been observed experimentally \cite{2023SY_HNL}.
  In Figure \ref{fig.xray} we show in addition to 2$\uptheta$--$\uptheta$ XRD spectrum,
  the intensity of the XRD data as a function of the growth conditions.  
}
As reported in Ref. \cite{2020JMO_HNLa}, the high-quality PdCrO$_2$ films can be achieved only
within a relatively narrow growth window.
Outside the growth window, the metallic properties are severely deteriorated by the impurity formation. 
The resistance could not be measured because of the high resistivity.
The rectangular boxes in Figure \ref{fig.xray} highlight the main impurities observed in XRD: Cr$_3$O$_4$ and Cr$_2$O$_3$.
The bottom two panels of Figure \ref{fig.xray} map the XRD intensities of Cr$_3$O$_4$ and Cr$_2$O$_3$ for temperature
and oxygen partial pressure.
\tadd{The relative abundances between Cr$_3$O$_4$ and Cr$_2$O$_3$ are difficult to assess quantitatively using  the XRD intensities }
\tadd{
  because the XRD reflectivity varies with substances and angles. However, we use the intensities
  to assess qualitatively how the formation of each substance is affected by growth conditions. 
}
The XRD intensity of Cr$_3$O$_4$ strongly positively correlates with temperature
(correlation coefficient
\footnote{
The correlation coefficient $r$ indicates the strength of correlation between two data-series, 
$\left\{X_i\right\}_{i=1}^{n}$ and $\left\{Y_i\right\}_{i=1}^{n}$. 
$r$ can be from -1 to +1. 
$r$ = +1 (-1) indicates the complete positive (negative) correlation 
between $\left\{X_i\right\}_{i=1}^{n}$ and $\left\{Y_i\right\}_{i=1}^{n}$. 
The correlation coefficient is defined by 
\begin{equation}
 r = \frac{\upsigma_{X,Y}}{\upsigma_{X}\upsigma_{Y}}. 
\end{equation}
Here, $\upsigma_{X}$ and $\upsigma_{Y}$ are the standard deviations of 
$\left\{X_i\right\}_{i=1}^{n}$ and $\left\{X_i\right\}_{i=1}^{n}$.  
$\upsigma_{X,Y}$ is the covariance of the two data-series: 
\begin{equation}
 \upsigma_{XY} = \frac{1}{n}\sum_{i=1}^{n}(X_i - \bar X)(Y_i-\bar Y).
\end{equation}
}
 $\uprho=+0.82$),
whereas the correlation between the XRD intensity of Cr$_2$O$_3$ and temperature is weak ($\uprho=+0.19$).
Moreover, Cr$_3$O$_4$'s peak strength does not depend on oxygen partial pressure ($\uprho=+0.01$), 
but Cr$_2$O$_3$'s peak strength negatively depends on oxygen partial pressure ($\uprho=-0.48$).
These results are \tadd{compared with our calculations} in the last paragraph of Section \ref{sec.entropy}. 

%%----------------------------------------
%\begin{figure}[htbp]
%  \centering
%  \includegraphics[width=\hsize]{str.eps} 
%  \caption{\label{fig.str}\ghost{fig.str}
%    High-resolution HAADF STEM image of an approximately 6 nm PdCrO$_2$ film
%    grown on an approximately 17 nm thick CuCrO$_2$ buffer layer.
%    The layer widths are different from the sample used for the present work's XRD experiments. This figure is quoted from unpublished work \cite{2023SY_HNL}
%  }
%\end{figure}
%%----------------------------------------
%----------------------------------------
%----------------------------------------

%%%%%%%%%%%%%%%%%%%%%%%%%%%%%%%%%%%%%%%%%
\subsection{Point-defect formation energy}
\label{sec.defect}\ghost{sec.defect}
%%%%%%%%%%%%%%%%%%%%%%%%%%%%%%%%%%%%%%%%%
Because the system segregates oxygen-deficient oxides Cr$_2$O$_3$ or Cr$_3$O$_4$, the segregation process should be accompanied by the appearance or disappearance of point defects. The ratio of O to Cr (O/Cr) in Cr$_2$O$_3$ is $\mathrm{O}/\mathrm{Cr}=1.5$. In Cr$_3$O$_4$, $\mathrm{O}/\mathrm{Cr}$ is about $1.3$. In the delafossite materials, $\mathrm{O}/\mathrm{Cr}$ is 2.0.
Therefore, the formation energies of multiple types of point self-defects
in bulk Al$_2$O$_3$, CuCrO$_2$, and PdCrO$_2$ were calculated.
To simplify the problem, defects in the bulk were calculated, even though samples with different 
thickness of PdCrO$_2$ and CuCrO$_2$ films have been grown in this and other works
\cite{2023SY_HNL,2020JMO_HNLa}.

\vspace{2mm}
The following describes the notation used for the point defects and explains how to evaluate the formation energies. 
$\mathrm{V_{Cr},V_{Cu},V_{Pd}, and \,V_{O}}$ indicate vacancies in the Cr, Cu, Pd, and O sites.
The Cu (Pd) replacement defects in the Cr sites, or \textit{antisite defects}, are indicated by Cu$_{\mathrm{Cr}}$ (Pd$_{\mathrm{Cr}}$).
Larger defect complexes such as Cu$_{\mathrm{Cr}}$\&$\mathrm{V_{Cu}}$ (Pd$_{\mathrm{Cr}}$\&$\mathrm{V_{Pd}}$) 
can form when Cu (Pd) atoms move to a preformed $\mathrm{V_{Cr}}$, leaving the $\mathrm{V_{Cr\,(Pd)}}$. 
The formation energies of these defects are given as follows:
for CuCrO$_2$,
%---------------
\begin{eqnarray}
  \Delta E\left( \mathrm{V_{\upalpha}} \right) &=& E\left( \mathrm{CuCrO_2} \right)_\mathrm{V_{\upalpha}} - E\left( \mathrm{CuCrO_2} \right)_\mathrm{bulk}
  \nonumber \\
  &+&  \upmu_{\mathrm{\upalpha}},\;\;\left(\upalpha=\mathrm{Cu,\,Cr,\,or\,O}\right),
  \\
  \Delta E\left( \mathrm{Cu_{Cr}\&V_{Cu}} \right)&=& E\left( \mathrm{CuCrO_2} \right)_\mathrm{Cu_{Cr}\&V_{Cu}} - E\left( \mathrm{CuCrO_2} \right)_\mathrm{bulk}
  \nonumber \\
  &+& \upmu_{\mathrm{Cr}},
\end{eqnarray}
%---------------
and for PdCrO$_2$,  
%---------------
\begin{eqnarray}
  \Delta E\left( \mathrm{V_{\upalpha}} \right)&=& E\left( \mathrm{PdCrO_2} \right)_\mathrm{V_{\upalpha}} - E\left( \mathrm{PdCrO_2} \right)_\mathrm{bulk}
  \nonumber \\
  &+& \upmu_{\mathrm{\upalpha}},\;\;\left(\upalpha=\mathrm{Pd,\,Cr\,or\,O}\right),
  \\
  \Delta E\left( \mathrm{Pd_{Cr}\&V_{Pd}} \right)&=& E\left( \mathrm{PdCrO_2} \right)_\mathrm{Pd_{Cr}\&V_{Pd}} - E\left( \mathrm{PdCrO_2} \right)_\mathrm{bulk}
  \nonumber \\
  &+& \upmu_{\mathrm{Cr}}. 
\end{eqnarray}
%---------------
Here, $E$(CuCrO$_2$)$_\mathrm{{bulk}}$ and $E$(PdCrO$_2$)$_\mathrm{{bulk}}$ are the total energies of pristine delafossite structures. 
$E$(CuCrO$_2$)$_X$ and $E$(PdCrO$_2$)$_X$ are the total energies of structures with type-$X$ defects. 
The chemical potential of atomic species $\upalpha$ is $\upmu_\upalpha$. 
The oxygen vacancy formation energy in the Al$_2$O$_3$ substrate was also calculated by 
\begin{eqnarray}
  \Delta E\left( \mathrm{V_O} \right) &=& E\left( \mathrm{Al_{2}O_{3}} \right)_{\mathrm{V_O}} - E\left( \mathrm{Al_{2}O_{3}} \right)_{\mathrm{bulk}} + \upmu_{\mathrm{O}}. 
\end{eqnarray}

\vspace{2mm}
\tadd{
  Our experiments observed the Cu-Pd alloy and Cr oxide impurity phases
  on the composite sample of Al$_2$O$_3$, CuCrO$_2$, and PdCrO$_2$ 
  (see \S~\ref{sec.segre}).
  The defect formation energies should be evaluated for 
  the experimental conditions: the chemical equilibrium states 
  consisting of Al$_2$O$_3$, CuCrO$_2$, PdCrO$_2$, Cu$_{x}$Pd$_{1-x}$, 
  and a chromium oxide.
}
The exact value of $x$ in Cu$_{x}$Pd$_{1-x}$ is not known experimentally. 
\tadd{
  The ratio $x$ potentially depends on the volume comparison of CuCrO$_2$ and PdCrO$_2$. 
  However, the change in the results is negligible when $x$ changes from 0.5 to 0.25 or 0.75
  (variations of only 0.33~eV were observed), as described in the Supporting Information.
} 
Therefore, the results reported below assumed $x=0.5$.
Solving the following equations yields the chemical potentials.
For example, if Al$_2$O$_3$, CuCrO$_2$, PdCrO$_2$, CuPd, and Cr$_3$O$_4$ coexist, then 
%---------------
\begin{eqnarray}
  2\upmu_\mathrm{Al}+3\upmu_\mathrm{O}&=&E\left(\mathrm{Al_2O_3}\right), \\
  \upmu_\mathrm{Cu}+\upmu_\mathrm{Cr}+2\upmu_\mathrm{O}&=&E\left(\mathrm{CuCrO_2}\right), \\
  \upmu_\mathrm{Pd}+\upmu_\mathrm{Cr}+2\upmu_\mathrm{O}&=&E\left(\mathrm{PdCrO_2}\right), \\
  \upmu_\mathrm{Cu}+\upmu_\mathrm{Pd}&=&E\left(\mathrm{CuPd}\right), \\
  3\upmu_\mathrm{Cr}+4\upmu_\mathrm{O}&=&E\left(\mathrm{Cr_3O_4}\right).
\end{eqnarray}
%---------------
There exist as many independent linear equations as unknown chemical potentials,
so the chemical potentials are trivially determined.
% \tadd{
%   Our experiments observed the Cu-Pd alloy and Cr oxide impurity phases
%   on the composite sample of Al$_2$O$_3$, CuCrO$_2$, and PdCrO$_2$ (see \S\ref{sec.segre}).
%   The defect formation energies should be evaluated for the experimental conditions:
%   the chemical equilibrium states consisting of Al$_2$O$_3$, CuCrO$_2$, PdCrO$_2$,
%   Cu$_{x}$Pd$_{1−x}$, and a chromium oxide.
% }

%%%%%%%%%%%%%%%%%%%%%%%%%%%%%%%%%%%%%%%%%
\subsection{Formation energies of defects as a function of the chemical potentials}
\begin{figure}[htbp]
  \centering
  \includegraphics[width=\hsize]{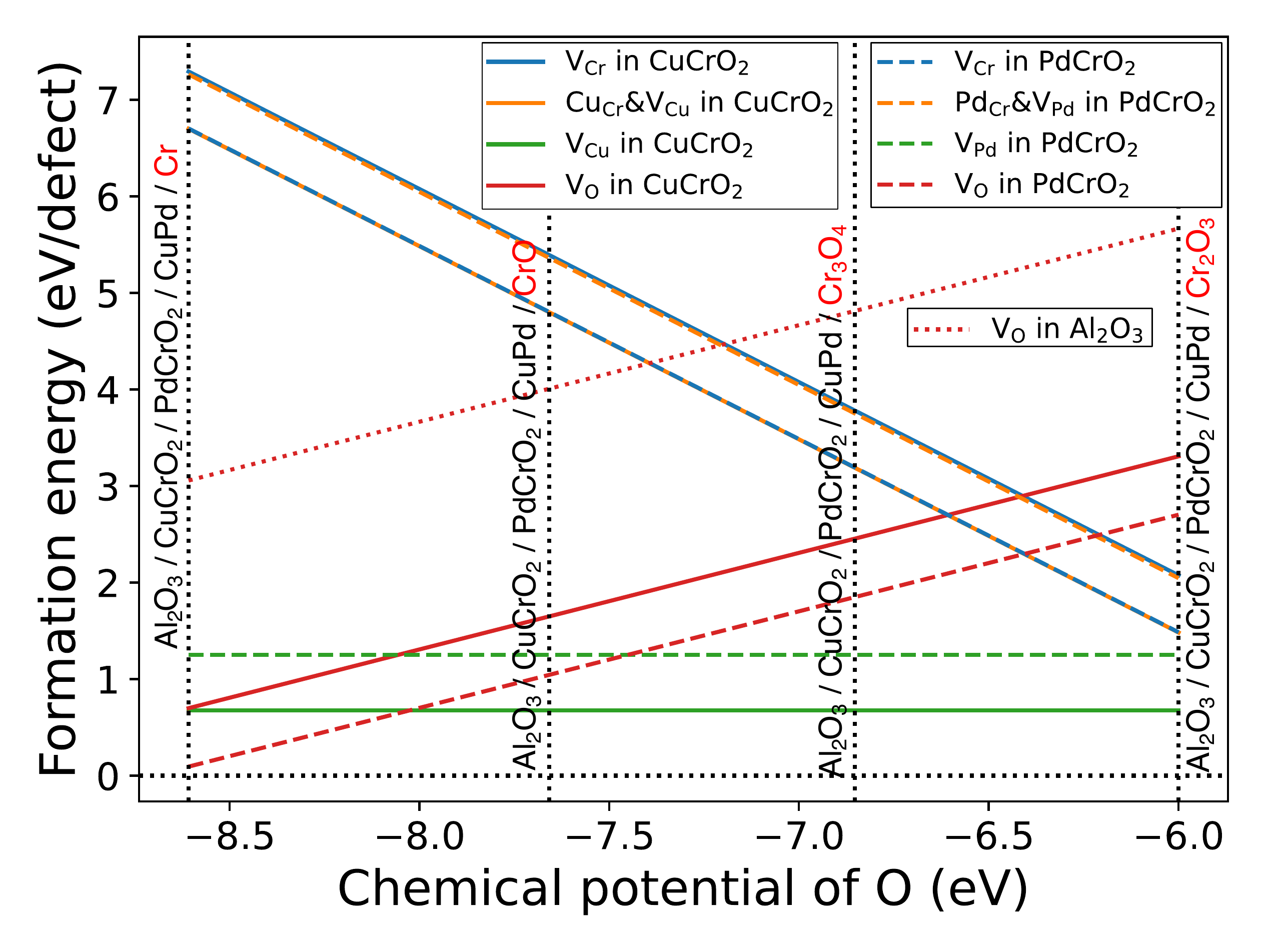}
  \caption{\label{fig.selfdef}\ghost{fig.selfdef}
    Formation energies of intrinsic point defects in CuCrO$_2$, PdCrO$_2$, and Al$_2$O$_3$
    calculated by the PBE+$U$ method as a function of the oxygen chemical potential.
    The chemical potentials are calculated for CuCrO$_2$, PdCrO$_2$, Al$_2$O$_3$, CuPd,
    and different chromium oxides.
}
\end{figure}

%%%%%%%%%%%%%%%%%%%%%%%%%%%%%%%%%%%%%%%%%
The formation energies of point defects in Al$_2$O$_3$, CuCrO$_2$, and PdCrO$_2$
were calculated for different chromium oxides, as described in Section \ref{sec.defect}.
\tadd{
  We also considered the Cr metal as the Cr source of the Cr-rich limit. 
}
The results are summarized in Figure \ref{fig.selfdef}.
The point-defect formation energies are all positive, so CuCrO$_2$ and PdCrO$_2$ are 
thermodynamically stable and stoichiometric under the considered chemical conditions. 
For low values of the oxygen chemical potential ($\lesssim -8.6$ eV), 
the $\mathrm{V_O}$ in CuCrO$_2$ and PdCrO$_2$ have the lowest formation energies;
the $\mathrm{V_O}$ in Al$_2$O$_3$ is much higher. 
As soon as the oxygen chemical potential increases,
the $\mathrm{V_{Cu}}$ and $\mathrm{V_{Pd}}$ become the lowest formation energy defects.
The Cr vacancies, by contrast, have much higher formation energies.
The experimental chemical potentials are not well defined 
because the system is out of equilibrium, as described in Section \ref{sec.defect}.
However, each element's stability corresponds to anywhere
between the vertical lines that correspond to the oxygen chemical potentials with $\mathrm{Cr_3O_4}$ and $\mathrm{Cr_2O_3}$.
The $\mathrm{V_{Cr}}$, $\mathrm{Cu_{Cr}}$\&$\mathrm{V_{Cu}}$, and $\mathrm{Pd_{Cr}}$\&$\mathrm{V_{Pd}}$ are all Cr-deficient point defects.
For CuCrO$_2$, the formation energy of $\mathrm{Cu_{Cr}}$\&$\mathrm{V_{Cu}}$ is lower than that of $\mathrm{V_{Cr}}$.
Therefore, the Cr site does not have a vacancy because a neighboring Cu occupies the Cr site by forming $\mathrm{V_{Cu}}$ 
next to $\mathrm{Cu_{Cr}}$.
%----------------------------------------
%-----------------------------------------

%%%%%%%%%%%%%%%%%%%%%%%%%%%%%%%%%%%%%%%%%%
\subsection{Instability of CuCrO$_2$ and PdCrO$_2$ for oxygen-deficient samples} 
%%%%%%%%%%%%%%%%%%%%%%%%%%%%%%%%%%%%%%%%%%
Experiments found the segregation of impurity phases of Cu$_{x}$Pd$_{1-x}$, Cr$_2$O$_3$, and Cr$_3$O$_4$
on a 10 nm PdCrO$_2$ layer with a one-monolayer CuCrO$_2$ buffer layer on an Al$_2$O$_3$ substrate. 
The samples were grown under low oxygen partial pressures.
The simultaneous presence of seven compunds (CuCrO$_2$, PdCrO$_2$,
Cu$_x$Pd$_{1-x}$, Cr$_2$O$_3$, Cr$_3$O$_4$, O$_2$, and Al$_2$O$_3$) but only five chemical elements
complicates the theoretical analysis.
Finding a solution for the chemical potential equations is impossible when the equations
outnumber the independent variables. In this case, the system is out of equilibrium. 
The chemical potentials may not be uniform throughuot the sample. 
For instance, near the surface, the oxygen chemical potential may be a function of temperature and oxygen partial pressure. 
By contrast, near the regions where Cr$_2$O$_3$ and Cr$_3$O$_4$ coexist, the chemical potentials of Cr and O are 
uniquely determined by the formation energies of the two solids.
Alternatively, near the Al$_2$O$_3$ substrate, 
the oxygen chemical potential may be determined by temperature and the concentration of oxygen vacancies in  Al$_2$O$_3$.
  
\vspace{2mm}
The impurity phases are oxygen deficient (i.e., chromium rich) relative to CuCrO$_2$ and PdCrO$_2$: 
the O/Cr ratios of Cr$_2$O$_3$ (1.5) and Cr$_3$O$_4$ ($\sim$1.3) are smaller than that of CuCrO$_2$ and PdCrO$_2$ (2.0). 
To elucidate the possible cause of impurity phase segregation, several different 
possible reactions originating from out-of-equilibrium states were considered. 
Then their potential to destabilize CuCrO$_2$ and PdCrO$_2$ was examined. 
The analysis revealed that low oxygen partial pressures and high temperatures could 
explain the segregation of Cr$_2$O$_3$ and Cu$_{x}$Pd$_{1-x}$. 
Preexisting defects as energetic as oxygen vacancies in Al$_2$O$_3$ could enhance 
the segregation of Cr$_3$O$_4$ and Cu$_{x}$Pd$_{1-x}$.

%%%%%%%%%%%%%%%%%%%%%%%%%%%%%%%%%%%%%%%%%
\subsection{Thermochemical reactions}
%%%%%%%%%%%%%%%%%%%%%%%%%%%%%%%%%%%%%%%%%
To simplify the analysis, the Cu$_x$Pd$_{1-x}$ alloy is assumed to be CuPd, as described in the last paragraph of Section \ref{sec.defect}. 
The theoretical approach shows that the combination 
of CuCrO$_2$ and PdCrO$_2$ is stable
against the CrO$_2$ and CuPd impurity phase segregation, which is a stoichiometric process.
The thermochemical equation of this segregation is 
%----------------------------------------
\begin{eqnarray}
  E(\mathrm{CuCrO_2}) + E(\mathrm{PdCrO_2}) &=& E(\mathrm{CuPd}) + 2E(\mathrm{CrO_2} ) \nonumber \\
  &+& Q\left(\mathrm{CrO_2} \right), \label{eq.q1}
\end{eqnarray}
%----------------------------------------
where $Q\left(\mathrm{CrO_2}\right)$ is the energy gained, or lost if negative,
to form CrO$_2$.
The value of $Q\left(\mathrm{CrO_2}\right)$ was calculated to be $-1.102$~eV per two formula units of CrO$_2$, so this reaction is endothermic.

\vspace{2mm}
By contrast,
the Cr/O ratios of CuCrO$_2$ and PdCrO$_2$ vs. Cr$_2$O$_3$ or Cr$_3$O$_4$ are different.
  Therefore, the impurity phase segregation may be caused by an impurity-absorbing defect.
For Cr$_2$O$_3$+CuPd, 
the impurity phase segregation may be caused and promoted by 
an oxygen-adsorbent mechanism because the Cr/O ratios of CuCrO$_2$ and
PdCrO$_2$ (1/2) and Cr$_2$O$_3$ (2/3) are different. 
This oxygen deficiency may be the result of low environmental oxygen concentration relative to chromium from either (i) defective CuCrO$_2$, PdCrO$_2$, or Al$_2$O$_3$ or (ii) low oxygen content 
in the vacuum growth chamber \footnote{The samples were grown under $<1000$ mTorr oxygen partial pressure.}. 
For mechanism (i), preexisting V$_{\mathrm{O}}$ in CuCrO$_2$, PdCrO$_2$, or Al$_2$O$_3$
and formation of Cr-deficient defects such as V$_{\mathrm{Cr}}$ in CuCrO$_2$ or PdCrO$_2$ were considered
to keep the Cr/O ratios constant before and after the process
\footnote{
Preexisting point defects with extra Cr,
such as $\mathrm{Cr_{Pd}}$ and $\mathrm{Cr_{Cu}}$,
were not considered because their existence explains the appearance of Cr$_x$O$_{1-x}$
but not Cu$_{y}$Pd$_{1-y}$ without involving multiple types defects
to balance the chemical reaction.
}. 
  
\vspace{2mm}
Therefore, the energy gain obtained by the (dis)appearance of point defects in CuCrO$_2$, PdCrO$_2$, and Al$_2$O$_3$ was compared with the
release of oxygen molecules into the oxygen gas in the growth chamber. 
All these possibilities were considered as particle exchanges with a particle bath.

\vspace{2mm}
The energy cost of taking an atom ($\upalpha =$ O or Cr) from one of these particle baths is defined as 
%----------------------------------------
\begin{eqnarray}
  \upnu_{\upalpha} \equiv E(\mathrm{\mathrm{bath}})_{\mathrm{bulk}} - E(\mathrm{\mathrm{bath}})_{\upalpha}. 
  \label{eq.nu}
\end{eqnarray}
%----------------------------------------
Here, $E(\mathrm{\mathrm{bath}})_{\mathrm{bulk}}$ and $E(\mathrm{\mathrm{bath}})_{\alpha}$ 
are the energies of the particle bath without defects and with an $\upalpha=$ O or Cr vacancy, respectively 
\footnote{$\upnu_{\upalpha}$ is conceptually similar to the chemical potential of $\upalpha$, but 
chemical potentials are not defined for a nonequilibrium process.}. 

\vspace{2mm}
The thermochemical equations for the segregation of Cr$_2$O$_3$ 
when introducing O to or removing Cr from the particle bath are given as follows: 
%----------------------------------------
\begin{eqnarray}
  E(\mathrm{CuCrO_2}) &+& E(\mathrm{PdCrO_2})
 = E(\mathrm{CuPd}) + E(\mathrm{Cr_2O_3})
  \nonumber \\
  &+& \upnu_{\mathrm{O}} + Q\left(\mathrm{Cr_2O_3, V_O^{rem}}\right), 
  \label{eq.q2} 
\end{eqnarray}
%----------------------------------------
%----------------------------------------
\begin{eqnarray}
  E(\mathrm{CuCrO_2}) &+& E(\mathrm{PdCrO_2})
  = E(\mathrm{CuPd}) + (4/3) E(\mathrm{Cr_2O_3})
  \nonumber \\
  &-& (2/3) \upnu_{\mathrm{Cr}} + Q\left(\mathrm{Cr_2O_3, V_{Cr}^{int}}\right).
  \label{eq.q3}
\end{eqnarray}
%----------------------------------------
\ghost{eq.q2,eq.q3}
In equation (\ref{eq.q2}), the term $\upnu_{\mathrm{O}}$ takes into account the effect of 
removing an oxygen vacancy in the particle bath, and $-(2/3)\upnu_{\mathrm{Cr}}$ 
considers the effect of creating a fraction of Cr vacancies in the particle bath. 

\vspace{2mm}
Similarly, the segregation of Cr$_3$O$_4$ could be explained by the following reactions: 
%----------------------------------------
\begin{eqnarray}
  E(\mathrm{CuCrO_2}) &+& E(\mathrm{PdCrO_2})
  = E(\mathrm{CuPd}) + (2/3) E(\mathrm{Cr_3O_4})
  \nonumber \\
  &+& (4/3) \upnu_{\mathrm{O}} + Q\left(\mathrm{Cr_3O_4, V_O^{rem}}\right), 
  \label{eq.q4} 
\end{eqnarray}
%----------------------------------------
%----------------------------------------
\begin{eqnarray}
  E(\mathrm{CuCrO_2}) &+& E(\mathrm{PdCrO_2})
  = E(\mathrm{CuPd}) + E(\mathrm{Cr_3O_4})
  \nonumber \\
  &-& \upnu_{\mathrm{Cr}} + Q\left(\mathrm{Cr_3O_4, V_{Cr}^{int}}\right). 
  \label{eq.q5} 
\end{eqnarray}
%----------------------------------------
\ghost{eq.q4,eq.q5}
Their derivations are described in detail in Appendix A.

\vspace{2mm}
The exothermic energies, $Q$, are shown in eqs (\ref{eq.q2})--(\ref{eq.q5}), for different values 
of $\upnu_{\mathrm{O}}$ and $\upnu_{\mathrm{Cr}}$, depending on the particle baths in Table \ref{tab.q25}. 
The table shows that only $Q\left(\mathrm{Cr_2O_3, V_O^{rem}}\right)$ and $Q\left(\mathrm{Cr_3O_4, V_O^{rem}}\right)$ 
can be positive (i.e., exothermic reaction), whereas 
the reactions involving the formation of Cr-deficient defects are always endothermic. 
Therefore, the preexisting oxygen vacancies could explain the 
spontaneous segregation of Cr$_2$O$_3$, Cr$_3$O$_4$, and CuPd impurity phases.

\begin{figure}[htbp]
  \centering
  \includegraphics[width=\hsize]{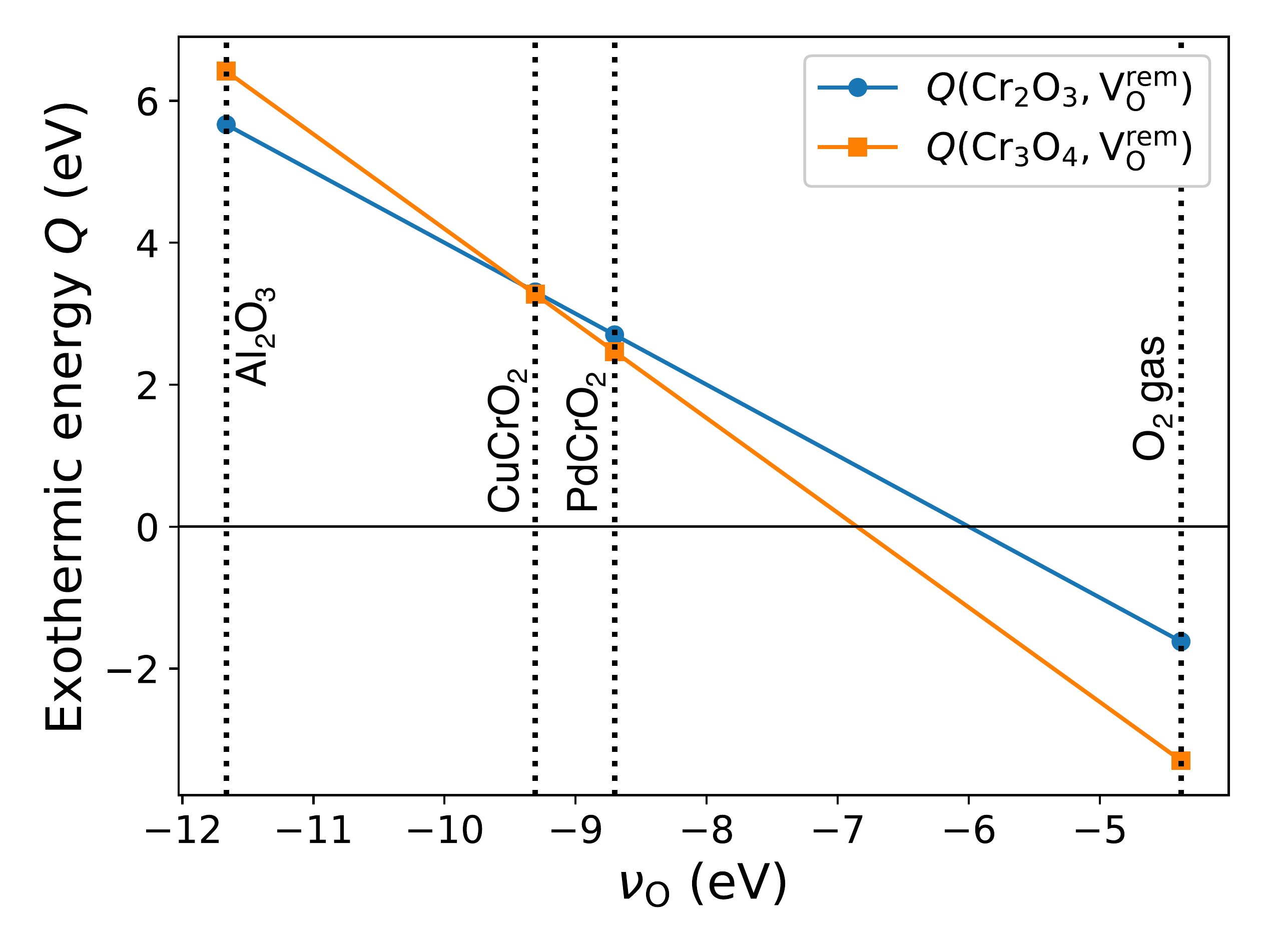} 
  \caption{\label{fig.zerok}\ghost{fig.zerok}
    Plots of $Q\left(\mathrm{Cr_2O_3, V_{O}^{rem}}\right)$ and $Q\left(\mathrm{Cr_3O_4, V_{O}^{rem}}\right)$ 
    shown in Table \ref{tab.q25} for different $\upnu_{\mathrm{O}}$ (i.e., different \tadd{O} particle baths). 
  }
\end{figure}

\vspace{2mm}
Figure \ref{fig.zerok} shows $Q\left(\mathrm{Cr_2O_3, V_{O}^{rem}}\right)$ and
$Q\left(\mathrm{Cr_3O_4, V_{O}^{rem}}\right)$ in Table \ref{tab.q25}
for different $\upnu_{\mathrm{O}}$ (i.e., different particle bath).
The stability of oxygen atoms in each particle bath negatively correlates with $\upnu_{\mathrm{O}}$:
oxygen atoms are the most (least) stable in $\mathrm{Al_2O_3}$ (O$_2$ gas).
$Q\left(\mathrm{Cr_2O_3, V_{O}^{rem}}\right)$ changes depending on $-\upnu_{\mathrm{O}}$, as given in eq (\ref{eq.q2}).
Similarly, $Q\left(\mathrm{Cr_3O_4, V_{O}^{rem}}\right)$ changes depending on $-(4/3)\upnu_{\mathrm{O}}$,
as given in eq (\ref{eq.q4}).
The impurity phase segregation is endothermic when O$_2$ gas is the particle bath 
and exothermic for the other particle baths.
The energetically favored chromium oxide changes from $\mathrm{Cr_2O_3}$ to $\mathrm{Cr_3O_4}$
with $\nu_{\mathrm{O}}$ decreasing from $\mathrm{PdCrO_2}$ to $\mathrm{Al_2O_3}$.

%----------------------------------------
\begin{table*}[htbp]
  \begin{center}
    \caption{\label{tab.q25}\ghost{tab.q25}
        Exothermic energies, $Q$ in eqs (\ref{eq.q2})--(\ref{eq.q5}),
        for the formation of Cr$_2$O$_3$ or Cr$_3$O$_4$ and CuPd accompanied 
        by $\mathrm{V_{O}^{rem}}$'s removal from or $\mathrm{V_{Cr}^{int}}$'s introduction to  
        different bath locations. 
    }
    \begin{tabular}{|c||cc|cc|}
      \hline  
      Bath location &
      $Q\left(\mathrm{Cr_2O_3, V_{O}^{rem}}\right)$ & $Q\left(\mathrm{Cr_2O_3, V_{Cr}^{int}}\right)$ &
      $Q\left(\mathrm{Cr_3O_4, V_{O}^{rem}}\right)$ & $Q\left(\mathrm{Cr_3O_4, V_{Cr}^{int}}\right)$ \\
      \hline
      PdCrO$_2$ & \textbf{+2.703} & $-$0.990 & \textbf{+2.466} & $-$3.192 \\
      CuCrO$_2$ & \textbf{+3.308} & $-$0.989 & \textbf{+3.273} & $-$3.189 \\
      Al$_2$O$_3$ & \textbf{+5.666} & -- & \textbf{+6.417} & -- \\
      O$_2$ in vacuum ($T=0$)   & $-$1.619 & -- & $–$3.297 & -- \\
      \hline
    \end{tabular}
  \end{center}
\end{table*}
%----------------------------------------
%----------------------------------------
%----------------------------------------

%%%%%%%%%%%%%%%%%%%%%%%%%%%%%%%%%%%%%%%%%
\subsection{Entropy contributions to the formation of\\Cr$_2$O$_3$ and Cr$_3$O$_4$}
\label{sec.entropy}\ghost{sec.entropy}
%%%%%%%%%%%%%%%%%%%%%%%%%%%%%%%%%%%%%%%%%
In this section, entropy contributions to the positive $Q\left(\mathrm{Cr_2O_3, V_O^{rem}}\right)$ 
and $Q\left(\mathrm{Cr_3O_4, V_O^{rem}}\right)$ are considered.
For convenience, $Q(\mathrm{Cr_2O_3}) \equiv Q\left(\mathrm{Cr_2O_3, V_O^{rem}}\right)$
and $Q(\mathrm{Cr_3O_4}) \equiv Q\left(\mathrm{Cr_3O_4, V_O^{rem}}\right)$.

\vspace{2mm}
Entropy contributions depend on the temperature, point-defect densities, and oxygen partial pressure. 
The energies $E$ were replaced by Helmholtz free energies $F(T)$ in eqs (\ref{eq.q2}) and (\ref{eq.q4}).
For bulk structures, $F(T)$ was evaluated by 
%----------------------------------------
\begin{eqnarray}
  F(T) = E + F_{\mathrm{vib}}(T).
\end{eqnarray}
%----------------------------------------
Here, $F_{\mathrm{vib}}(T)$ is the vibrational free energy. 
For $\upnu_\mathrm{O}$ in a defective solid, the vacancy configurational entropy 
contribution was considered in addition to $F_{\mathrm{vib}}(T)$. 
If the vacancy density is $\mathrm{c_v}$, then the free energy change when removing one vacancy is 
%----------------------------------------
\begin{eqnarray}
  \Delta F_{\mathrm{config}}(T,c_\mathrm{v})= k_{\mathrm{B}}T[-\ln{(c_\mathrm{v})}+\ln{(1-c_\mathrm{v})}] 
\end{eqnarray}
%----------------------------------------
(details in Appendix B).
Here, $k_{\mathrm{B}}$ is the Boltzmann constant. 
Therefore, $Q(\mathrm{Cr_2O_3})$ and $Q(\mathrm{Cr_3O_4})$ depend on vacancy density and temperature
when the bath location is a defective solid. 

\vspace{2mm}
When considering the case of O$_2$ released into the growth chamber, because the experimental oxygen partial pressure is very low and the temperature is high, 
the translational entropy contribution could significantly stabilize the oxygen gas. 
This stabilization may change $Q(\mathrm{Cr_2O_3})$ and $Q(\mathrm{Cr_3O_4})$ from negative to positive.
Without entropy contributions, they are negative, as shown in Table \ref{tab.q25}.
The Helmholtz free energy $F(T)$ of the oxygen gas per molecule is defined as 
%----------------------------------------
\begin{eqnarray}
  F(T,P_{\mathrm{O_2}})=E(\mathrm{O_2})+F_\mathrm{vib}(T)+F_\mathrm{rot}(T)+F_\mathrm{trans}(T,P_{\mathrm{O_2}})
\end{eqnarray}
%----------------------------------------
Here, $E(\mathrm{O_2})$ is the energy of an isolated oxygen molecule and
$F_\mathrm{vib}(T)$, $F_\mathrm{rot}(T)$, and $F_\mathrm{trans}(T,P_{\mathrm{O_2}})$ are
free energies by vibrational, rotational, and translational entropies, respectively.
Then $F_\mathrm{rot}(T)$ and $F_\mathrm{trans}(T,P_{\mathrm{O_2}})$ \cite{1985DLG} are given by 
%----------------------------------------
\begin{eqnarray}
  F_{\mathrm{rot}}(T)  &=& -k_{\mathrm{B}}T\left( 1+\ln{\frac{8\pi^2Ik_\mathrm{B}T}{2h^2}} \right), \\
  F_{\mathrm{trans}}(T,P_{\mathrm{O_2}})  &=& -k_{\mathrm{B}}T\ln{\frac{k_{\mathrm{B}}T}{P_{\mathrm{O_2}}\Lambda^3}}, \\
  \Lambda &\equiv& \frac{h}{\sqrt{2 \pi m k_{\mathrm{B}} T}}
\end{eqnarray}
%----------------------------------------
Here, $h$ is the Planck constant, and 
$I$ is the moment of inertia of an oxygen molecule.
Therefore, $Q(\mathrm{Cr_2O_3})$ and $Q(\mathrm{Cr_3O_4})$ depend on oxygen partial pressure and temperature
when the bath location is the dilute oxygen gas.

\vspace{2mm}
The entropy contributions in eqs (\ref{eq.q2}) and (\ref{eq.q4}) 
yield $Q(\mathrm{Cr_2O_3})$ and $Q(\mathrm{Cr_3O_4})$ for different bath locations and conditions. 
In these equations, the bulk free energies depend on only the temperature. 
When the bath is a defected crystal, $\upnu_{\mathrm{O}}$ depends on temperature and vacancy density. 
When the bath is the oxygen gas, $\upnu_{\mathrm{O}}$ depends on temperature and oxygen partial pressure. 
When $\upnu_{\mathrm{O}}$'s entropy contributions are ignored,  $Q(\mathrm{Cr_2O_3})$ and $Q(\mathrm{Cr_3O_4})$ 
barely depend on the temperature: 
$Q(\mathrm{Cr_2O_3})$ and $Q(\mathrm{Cr_3O_4})$ do not change more than 60 meV from 600 to 1000 K,
and $Q(\mathrm{Cr_2O_3}) - Q(\mathrm{Cr_3O_4})$ does not change more than 11 meV from 600 to 1000 K. 
Therefore, the dependence of $Q(\mathrm{Cr_2O_3})$ and $Q(\mathrm{Cr_3O_4})$ on the conditions 
is almost equivalent to that of $\upnu_{\mathrm{O}}$.

\begin{figure}[htbp]
  \centering
  \includegraphics[width=\hsize]{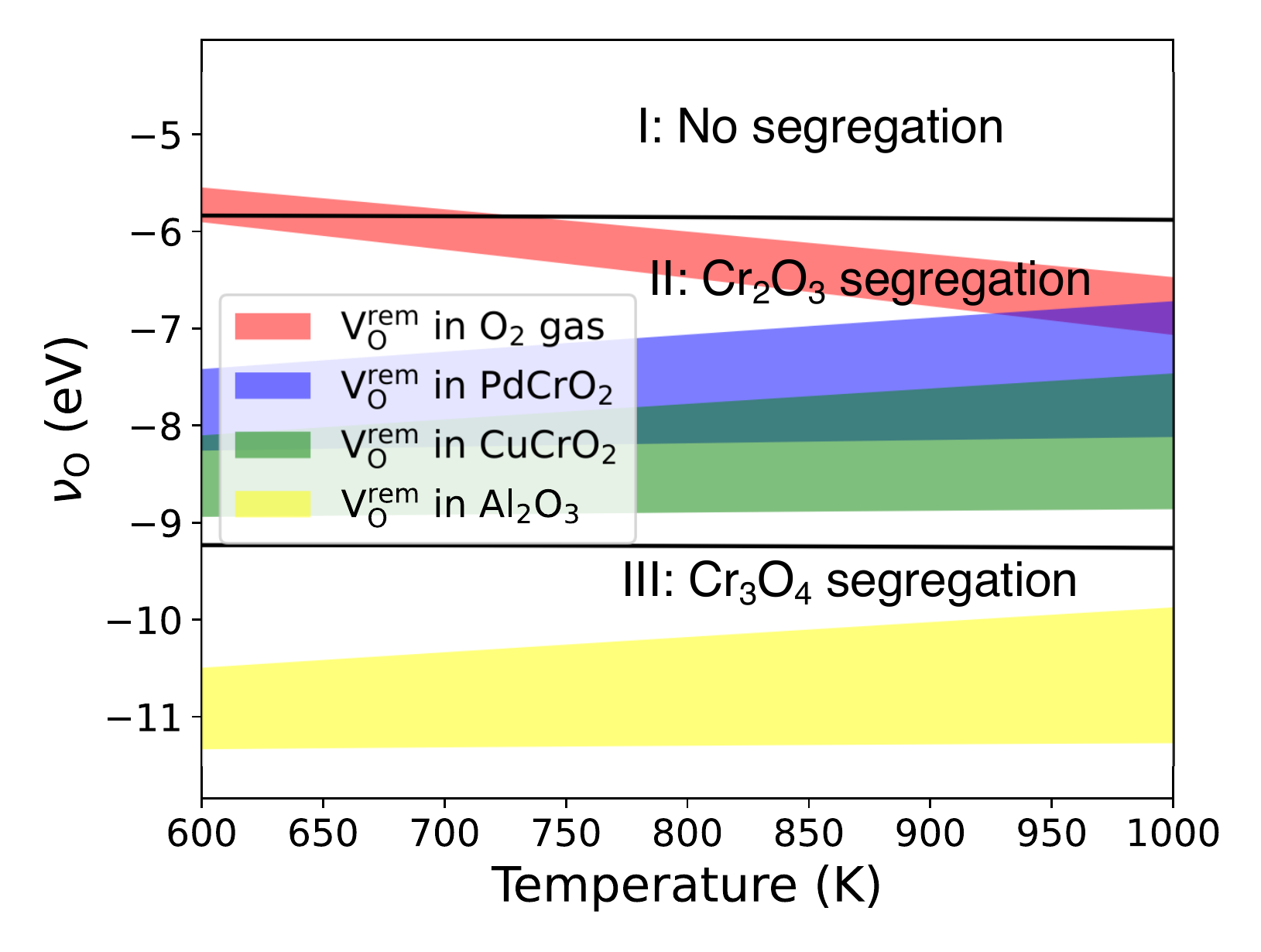}
  \caption{\label{fig.nuplot}\ghost{fig.nuplot}
    Plots of the energy of the oxygen sink ({\nuoxy}) using eq (\ref{eq.nu}) for different possible locations in the 
    experimental range of temperatures and estimated concentration.
     The upper and lower edges of $\mathrm{V_{O}^{rem}}$ in O$_2$ gas are $P_{\mathrm{O_2}}$ = 1 and 10$^{-6}$ atm.
     The upper and lower edges of the other areas are $c_\mathrm{v}$ = $10^{-1}$ and $10^{-8}$ per site.
    The dark green area between the blue and green areas
    is the overlap of the blue and green areas. 
  }
\end{figure}

\vspace{2mm}
To understand the conditions under which different oxides might be generated experimentally,
different baths for exchanging oxygen were systematically considered.
When the bath is a defected crystal, $\upnu_\mathrm{O}$ was calculated for vacancy densities in the range of $10^{-8}$--$10^{-1}$ per site and temperatures in the range of 600--1000 K. 
When the bath is the oxygen gas, $\upnu_\mathrm{O}$ was calculated for oxygen partial pressures in the range of $10^{-6}$--$10^{0}$ atm and temperatures in the range of 600--1000 K. 
Figure \ref{fig.nuplot} shows a map of the $\upnu_{\mathrm{O}}$ calculated for different bath locations. 
The vertical width of each area indicates the variation width corresponding to vacancy densities in the range of 
$10^{-8}$--$10^{-1}$ per site or oxygen partial pressures in the range of $10^{-6}$--$10^{0}$ atm.
Figure \ref{fig.nuplot} is divided into the three regions I--III 
according to the corresponding $Q(\mathrm{Cr_2O_3})$ and $Q(\mathrm{Cr_3O_4})$ values. 
The region I is $Q(\mathrm{Cr_2O_3, Cr_3O_4}) < 0$: the impurity phase segregation does not proceed spontaneously. 
The region II is $Q(\mathrm{Cr_2O_3})>0$ and $Q(\mathrm{Cr_2O_3})>Q(\mathrm{Cr_3O_4})$: 
$\mathrm{Cr_2O_3 + Cu_xPd_{1-x}}$ is spontaneously predominantly formed. 
The region III is $Q(\mathrm{Cr_3O_4})>Q(\mathrm{Cr_2O_3})>0$: 
$\mathrm{Cr_3O_4 + Cu_xPd_{1-x}}$ is spontaneously predominantly formed. 
Therefore, when the particle bath is oxygen gas, CuCrO$_2$, or PdCrO$_2$, 
the majority of chromium oxide is $\mathrm{Cr_2O_3}$. 
When the particle bath is $\mathrm{Al_2O_3}$, the majority of chromium oxide is $\mathrm{Cr_3O_4}$.

\vspace{2mm}
Furthermore, other bath locations than the above listed are realistically possible. 
For example, an oxygen-terminated PdCrO$_2$ surface could lead to very high $\upnu_\mathrm{O}$. 
By contrast, a Pd-terminated surface could lead to very low $\upnu_\mathrm{O}$. 
Investigation of such further complicated mechanisms is a possible future work for theory and experiments.

\vspace{2mm}
These calculations revealed that $\upnu_{\mathrm{O}}$ in O$_2$ gas decreases with decreasing oxygen partial pressure, 
and $\upnu_{\mathrm{O}}$ in defected crystals decreases with increasing $c_{\mathrm{v}}$ (details in supporting information).  
This result is not surprising because it indicates that gaseous oxygen molecules are more stable under oxygen-poor conditions. 
In reality, $c_\mathrm{v}$ would negatively correlates with oxygen partial pressure, 
so $\upnu_{\mathrm{O}}$ positively correlates with oxygen partial pressure in every bath location: 
lower oxygen partial pressure facilitates the segregation of impurity phases.
This analysis agrees with the experimental finding described in Section \ref{sec.segre}: 
$\mathrm{Cr_2O_3}$ formation negatively correlates with oxygen partial pressure. 

\vspace{2mm}
However, this analysis does not explain the independence of $\mathrm{Cr_3O_4}$ formation on oxygen partial pressure. 
Rather, $\mathrm{Cr_3O_4}$ formation strongly depends on temperature, unlike $\mathrm{Cr_2O_3}$. 
Some hypotheses are considered to explain the $\mathrm{Cr_3O_4}$ experimental results. 
(i) Most of the bath locations belong to region II in Figure \ref{fig.nuplot},
and the temperature determines how much the metastable Cr$_3$O$_4$ is segregated. 
(ii) Some of the bath locations belong to region III, but high barrier energy is required
to transfer oxygen atoms, moving oxygen vacancies, so the temperature determines the oxygen exchange rate.

\vspace{2mm}
For example, the barrier energy required to transfer oxygen atoms from Al$_2$O$_3$ to the surface 
would be higher than from PdCrO$_2$ to the surface. 
To verify the hypotheses, saddle state analyses by methods such as the nudged elastic band method, molecular dynamics,
and/or modeling the sample's surface should be applied in future works.   

%----------------------------------------
%----------------------------------------

%%%%%%%%%%%%%%%%%%%%%%%%%%%%%%%%%%%%%%%%%
\section{Conclusion}
\label{sec.conclusion}\ghost{sec.conclusion}
%%%%%%%%%%%%%%%%%%%%%%%%%%%%%%%%%%%%%%%%%
The mechanism of impurity phase segregation with the epitaxial growth 
of a PdCrO$_2$ layer on a CuCrO$_2$ buffer layer on an Al$_2$O$_3$ substrate 
was investigated via a combination of experiments and ab initio calculations. 
XRD experiments revealed the formation of Cu$_{x}$Pd$_{1-x}$ alloy 
and chromium oxide (Cr$_2$O$_3$ and Cr$_3$O$_4$) impurity phases. 
Consequently, the impurity phase segregation should be 
involved with appearance or disappearance of point defects or oxygen migration 
because the possible segregation processes are not stoichiometric. 
In this scenario, several possible mechanisms of impurity phase segregation were considered 
with oxygen vacancy disappearance or chromium vacancy appearance into different particle baths: 
Al$_2$O$_3$, CuCrO$_2$, PdCrO$_2$, and the dilute oxygen gas. 
Calculations established that the oxygen vacancy consumption processes are energetically favorable
and supported experimental evidence that Cr$_2$O$_3$ or Cr$_3$O$_4$ are the predominant chromium oxide impurity phases.
Specifically, preventing the release of oxygen atoms from delafossite materials
could suppress the impurity phase segregation. 

%%%%%%%%%%%%%%%%%%%%%%%%%%%%%%%%%%%%%%%%%
\section*{Appendix}
%%%%%%%%%%%%%%%%%%%%%%%%%%%%%%%%%%%%%%%%%
\subsection{Derivation of eqs (\ref{eq.q2}) and (\ref{eq.q3})}
\label{app.derive}\ghost{app.derive}
%%%%%%%%%%%%%%%%%%%%%%%%%%%%%%%%%%%%%%%%%
For ease of explanation, let the particle bath be PdCrO$_2$.
Consider the following thermochemical equations for the segregation of Cr$_2$O$_3$
by removing preexisting O vacancies or creating Cr vacancies.
%----------------------------------------
\begin{eqnarray}    
  && E(\mathrm{CuCrO_2})_{\mathrm{bulk}}^{(n)} + E(\mathrm{PdCrO_2})_{\mathrm{V_{O}}}^{(n)}
  \nonumber \\
  &=& E(\mathrm{CuCrO_2})_{\mathrm{bulk}}^{(n-1)} + E(\mathrm{PdCrO_2})_{\mathrm{bulk}}^{(n-1)}
  \nonumber \\
  &+& E(\mathrm{CuPd}) + E(\mathrm{Cr_2O_3}) + Q\left(\mathrm{Cr_2O_3, V_O^{rem}}\right) \label{eq.q2pre}     
\end{eqnarray}  
%----------------------------------------
and 
%----------------------------------------
\begin{eqnarray}    
  && E(\mathrm{CuCrO_2})_{\mathrm{bulk}}^{(n)}
  + E(\mathrm{PdCrO_2})_{\mathrm{bulk}}^{(n)}
  \nonumber \\
  &=& E(\mathrm{CuCrO_2})_{\mathrm{bulk}}^{(n-1)} + E(\mathrm{PdCrO_2})_{\mathrm{(2/3)V_{Cr}}}^{(n-1)}
  \nonumber \\
  &+& E(\mathrm{CuPd}) + (4/3)E(\mathrm{Cr_2O_3}) + Q\left(\mathrm{Cr_2O_3, V_{Cr}^{int}}\right). \label{eq.q3pre}     
\end{eqnarray}  
%----------------------------------------
Here, for example, $E(\mathrm{CuCrO_2})_{\mathrm{bulk}}^{(n)}$ is the energy of $n$ f.u. bulk CuCrO$_2$, 
$E(\mathrm{PdCrO_2})_{\mathrm{V_{O}}}^{(n)}$ is the energy of $n$ f.u. PdCrO$_2$ with an oxygen vacancy,
$E(\mathrm{PdCrO_2})_{(2/3)\mathrm{V_{Cr}}}^{(n-1)}$ is the energy of $(n-1)$ f.u. PdCrO$_2$ 
with a fraction of $2/3$ chromium vacancies, and $E(\mathrm{CuPd})$ is the energy of 1 f.u. bulk CuPd. 
For the bulk, the following relationships hold according to the definitions. 
%----------------------------------------
\begin{eqnarray}
  E(\mathrm{CuCrO_2})_{\mathrm{bulk}}^{(n)} &=& n \, E(\mathrm{CuCrO_2}), 
  \label{eq.supple1} \\
  E(\mathrm{PdCrO_2})_{\mathrm{bulk}}^{(n)} &=& n \, E(\mathrm{PdCrO_2}).
  \label{eq.supple2}
\end{eqnarray}
%----------------------------------------
Define the energy gain from removing $m$ oxygen or chromium vacancies 
in $n \mathrm{PdCrO_2}$ as follows: 
%----------------------------------------
\begin{eqnarray}
  \nu_{\mathrm{O}} (n,m) &\equiv& E(\mathrm{PdCrO_2})_{\mathrm{bulk}}^{(n)} - E(\mathrm{PdCrO_2})_{m\,\mathrm{V_{O}}}^{(n)},
  \label{eq.supple3} \\
  \upnu_{\mathrm{Cr}}(n,m) &\equiv& E(\mathrm{PdCrO_2})_{\mathrm{bulk}}^{(n)} - E(\mathrm{PdCrO_2})_{m\,\mathrm{V_{Cr}}}^{(n)}.
  \label{eq.supple4}
\end{eqnarray}  
%----------------------------------------
In the thermodynamic limit ($n \rightarrow \infty$), 
the following relationships should hold: 
%----------------------------------------
\begin{eqnarray}
  \nu_{\mathrm{O,Cr}} (n,m) &\simeq& \upnu_{\mathrm{O,Cr}} (n-1,m), 
  \label{eq.supple5} \\ 
  \nu_{\mathrm{O,Cr}} (n,m) &\simeq& m \, \upnu_{\mathrm{O,Cr}} (n,1),
  \label{eq.supple6} 
\end{eqnarray}
%----------------------------------------
Moreover, define 
%----------------------------------------
\begin{equation}
  \upnu_{\mathrm{O,Cr}} \equiv \upnu_{\mathrm{O,Cr}} (n,1)
  \label{eq.supple7}
\end{equation}
%----------------------------------------
These are the $\upnu_{\upalpha}$ defined in eq (\ref{eq.nu}).
Applying eqs (\ref{eq.supple1})--(\ref{eq.supple7}) to eqs (\ref{eq.q2pre}) and (\ref{eq.q3pre}), yields eqs (\ref{eq.q2}) and (\ref{eq.q3}).

%%%%%%%%%%%%%%%%%%%%%%%%%%%%%%%%%%%%%%%%%
\subsection{Configurational entropy of removing a vacancy.}
\label{app.entropy}\ghost{app.entropy}
%%%%%%%%%%%%%%%%%%%%%%%%%%%%%%%%%%%%%%%%%
When $n$ vacancies exist in $N$ sites, the configurational entropy is 
\begin{eqnarray}
  S(N,n)=k_\mathrm{B}\ln{\frac{N!}{(N-n)!n!}}.
\end{eqnarray}
The entropy change achieved by adding one vacancy is given by
\begin{eqnarray}
  \Delta S(N,n) &\equiv& S(N,n+1) - S(N,n) \nonumber \\
  &=& k_\mathrm{B}\left[-\ln{(c_\mathrm{v}+1/N)}+\ln{(1-c_\mathrm{v})}\right], \\
  c_\mathrm{v} &\equiv& n/N. 
\end{eqnarray}
For the limit of $N \to \infty$ with fixed $c_v$,
\begin{eqnarray}
  \Delta S(N,n) \to \Delta S(c_\mathrm{v}) = k_\mathrm{B}\left[-\ln{(c_\mathrm{v})}+\ln{(1-c_\mathrm{v})}\right]
\end{eqnarray}
Therefore, the free energy change achieved by removing one vacancy is 
given by
\begin{eqnarray}
  \Delta F(T,c_v) &=& -T(-S(c_\mathrm{v})) \\
  &=& k_\mathrm{B}T\left[-\ln{(c_\mathrm{v})}+\ln{(1-c_\mathrm{v})}\right].
\end{eqnarray}

%%%%%%%%%%%%%%%%%%%%%%%%%%%%%%%%%%%%%%%%%
\section{Acknowledgments}
%%%%%%%%%%%%%%%%%%%%%%%%%%%%%%%%%%%%%%%%%
We acknowledge E. Heinrich for valuable help with manuscript preparation.
This work was supported by the US Department of Energy, Office of Science,
Basic Energy Sciences, Materials Sciences and Engineering Division
\tadd{(theory and synthesis) and as part of the Computational Materials Sciences Program 
and Center for Predictive Simulation of Functional Materials (structural characterization)}.
VESTA \cite{2011KM_FI} was used to draw the crystal structures. 

%%%%%%%%%%%%%%%%%%%%%%%%%%%%%%%%%%%%%%%%%
\bibliography{references}
\end{document}